\documentclass[fleqn,usenatbib]{mnras}

\usepackage{newtxtext,newtxmath}

\usepackage[T1]{fontenc}

\DeclareRobustCommand{\VAN}[3]{#2}
\let\VANthebibliography\thebibliography
\def\thebibliography{\DeclareRobustCommand{\VAN}[3]{##3}\VANthebibliography}

\usepackage{graphicx}	
\usepackage{amsmath}	

\title[50 years of PSR J0738$-$4042]{Rotational and radio emission properties of PSR~J0738$-$4042 over half a century}

\author[M. E. Lower et al.]{\parbox{\textwidth}
{
M.~E.~Lower,$^{1}$\thanks{E-mail: marcus.lower@csiro.au}
S.~Johnston,$^{1}$
A.~Karastergiou,$^{2,3}$
P.~R. Brook,$^{4}$
M.~Bailes,$^{5,6}$
S.~Buchner,$^{7}$
A.~T.~Deller,$^{5,6}$
L.~Dunn,$^{8,6}$
C.~Flynn,$^{5}$
M.~Kerr,$^{9}$
R.~N.~Manchester,$^{1}$
A.~Mandlik,$^{5}$
L.~S.~Oswald,$^{2,10}$
A.~Parthasarathy,$^{11}$
R.~M.~Shannon,$^{5,6}$
C.~Sobey$^{12,13}$
and P.~Weltevrede$^{14}$
}
\\ \\
$^{1}$Australia Telescope National Facility, CSIRO, Space and Astronomy, PO Box 76, Epping, NSW 1710, Australia\\
$^{2}$Department of Astrophysics, University of Oxford, Denys Wilkinson Building, Keble Road, Oxford OX1 3RH, UK\\
$^{3}$Department of Physics and Electronics, Rhodes University, PO Box 94, Grahamstown 6140, South Africa\\
$^{4}$Institute for Gravitational Wave Astronomy and School of Physics and Astronomy, University of Birmingham, Edgbaston, Birmingham B15 2TT, UK\\
$^{5}$Centre for Astrophysics and Supercomputing, Swinburne University of Technology, PO Box 218, Hawthorn, VIC 3122, Australia\\
$^{6}$OzGrav: The ARC Centre of Excellence for Gravitational-wave Discovery\\
$^{7}$South African Radio Astronomy Observatory (SARAO), 2 Fir Street, Black River Park, Observatory, Cape Town 7925, South Africa\\
$^{8}$School of Physics, University of Melbourne, Parkville, VIC 3010, Australia\\
$^{9}$Space Science Division, Naval Research Laboratory, Washington, DC 20375, USA\\
$^{10}$Magdalen College, University of Oxford, Oxford OX1 4AU, UK\\
$^{11}$Max-Planck-Institut f{\"u}r Radioastronomie, Auf dem H{\"u}gel 69, D-53121 Bonn, Germany\\
$^{12}$SKA Observatory, SKA-LOW Science Operations Centre, ARRC Building, 26 Dick Perry Avenue, Technology Park, Kensington, WA 6151, Australia\\
$^{13}$CSIRO, Space and Astronomy, PO Box 1130, Bentley, WA 6102, Australia\\
$^{14}$Jodrell Bank Centre for Astrophysics, The University of Manchester, Alan Turing Building, Manchester M13 9PL, United Kingdom
}

\date{Accepted XXXX. Received YYYY; in original form ZZZZ}

\begin{document}
\label{firstpage}
\pagerange{\pageref{firstpage}--\pageref{lastpage}}
\maketitle

\begin{abstract}
We present a comprehensive study of the rotational and emission properties of PSR~J0738$-$4042 using a combination of observations taken by the Deep Space Network, Hartebeesthoek, Parkes (Murriyang) and Molonglo observatories between 1972 and 2023.
Our timing of the pulsar is motivated by previously reported profile/spin-down events that occurred in September 2005 and December 2015, which result in an anomalously large braking index of $n = 23300 \pm 1800$.
Using a Gaussian process regression framework, we develop continuous models for the evolution of the pulsar spin-down rate ($\dot{\nu}$) and profile shape.
We find that the pulse profile variations are similar regardless of radio observing frequency and polarisation.
Small-scale differences can be ascribed to changes in the interstellar medium along the line of sight and frequency-dependent changes in magnetospheric radio emission height.
No new correlated spin-down or profile events were identified in our extended dataset.
However, we found that the disappearance of a bright emission component in the leading edge of archival profiles between 1981--1988 was not associated with a substantial change in $\dot{\nu}$.
This marks a notable departure from the previous profile/spin-down events in this pulsar.
We discuss the challenges these observations pose for physical models and conclude that interactions between the pulsar and in-falling asteroids or a form of magnetospheric state-switching with a long periodicity are plausible explanations.
\end{abstract}

\begin{keywords}
radiation mechanisms: non-thermal -- stars: neutron -- pulsars: individual: PSR~J0738$-$4042
\end{keywords}



\section{Introduction}

Pulsars are a class of rotating neutron star that emit coherent beams of radiation from regions around their magnetic poles.
They are detected as sources of periodic radio pulses when these beams sweep across the Earth. 
While individual radio pulses can exhibit substantial shape variations, it is often observed that the average profile shape remains stable over years to decades while the pulsar smoothly spins down as rotational angular momentum is lost to dipole radiation, particle winds and gamma-ray emission.  
However, there is a growing population of pulsars that are observed to switch between two or more quasi-stable emission states on timescales ranging between individual rotations of the neutron star to many months or years \citep{Hermsen2013, Hermsen2017, Hermsen2018, Stairs2019}.
Mode-switching predominantly manifests as changes in the pulse profile, including the relative intensity, rotation phase and widths of individual profile components, or in extreme cases, nulls where the radio emission either shuts off completely or falls below detectable levels before switching back on again \citep{Wang2007, Chen2020}.
It has been previously linked to coincident variations in the spin-down rates of several pulsars \citep{Kramer2006, Lyne2010, Brook2016, Shaw2022}.
Although the exact mechanism responsible for this phenomenon is presently unknown, the prevailing theory is that mode-switching is caused by large-scale changes in the electric currents and plasma properties of pulsar magnetospheres, where the corresponding change in particle wind density alters the torque acting to slow the pulsar rotation \citep{Timokhin2010}.
Other more speculative origins for mode-changing include variations in viewing or magnetic geometry \citep{Kou2018}, free precession \citep{Kerr2016}, starquakes \citep{Zhou2023}, and interactions with minor bodies or debris disks \citep{Cordes2008}.

Many pulsars display such stochastic irregularities in their rotation that appear as structures in their timing residuals.
Fluctuations that have steep spectra, i.e. dominant at  low fluctuation frequencies, are often referred to as `timing noise', a phenomenon that can also arise from the dynamics of neutron star interiors \citep{Melatos2014}.
The other main contributor to timing noise are glitches, as sudden spin-up events caused by either starquakes \citep{Ruderman1969, Baym1969} or an exchange of angular momentum from the superfluid core to the neutron star crust \citep{Anderson1975, Alpar1984a, Melatos2008}. 
Several studies over the past decade have suggested possible links between between glitches and profile variations in a small sample of mode-changing pulsars (\citealt{Weltevrede2011}\footnote{PSR~J1119$-$6127 is a high magnetic-field strength pulsar that occasionally displays magnetar-like behaviour (see \citealt{Weltevrede2011, Dai2018})}; \citealt{Keith2013, Kou2018, Takata2020, Ge2020, Zhou2023}).

One noteworthy pulsar that exhibits significant correlated variations in its radio profile and spin-down evolution is PSR~J0738$-$4042 (PSR~B0736$-$40).
Discovered during the first pulsar surveys of the Molonglo Radio Observatory  \citep{Large1968}, PSR~J0738$-$4042 is a bright (112\,mJy at 1.4 GHz) radio pulsar with a spin-frequency of $\nu = 2.667$\,Hz and average spin-down rate of about $\dot{\nu} = -1 \times 10^{-14}$\,s$^{-2}$ \citep{Jankowski2019}.
It is known to display significant changes in its average pulse profile shape over time. 
The most notable example of this behaviour was the emergence of a precursor component in the leading edge of its average pulse profile in 2005 \citep{Karastergiou2011}.
A follow-up study of high-cadence monitoring of PSR~J0738$-$4042 at the Hartebeesthoek Radio Astronomy Observatory (HartRAO) and Parkes Observatory by \citet{Brook2014} revealed the recently emerged component drifted by approximately $+7.2$\degr\ in pulse phase over $\sim$100\,days.
This behaviour coincided with an initial enhancement of the spin-down rate by $1.4 \times 10^{-15}$\,s$^{-2}$ over the same 100\,day period, followed by several smaller increases and decreases in $\dot{\nu}$ that eventually stabilised $\sim$1000\,days after the event. 
They suggested that both the rapid changes in spin-down and new profile component could have been caused by the pulsar interacting with an in-falling asteroid, where the ionized remains of the asteroid triggered a pair cascade in a previously inactive region of the magnetosphere \citep{Cordes2008}.
An increase in the precursor component intensity (with respect to the profile peak) was observed in 2010, however this event did not coincide with any change in the neutron star spin-down from the lower, post-2005 event value \citep{Brook2014, Brook2016}.
Note that throughout this work we will be referring to increases or deficits of profile-component emission intensity or amplitude relative to the profile peak, not the absolute flux density of the pulsar.
More recently, \citet{Zhou2023} discovered a sudden decrease in the precursor amplitude and increased intensity in the leading edge of the main component took place in late-2015.
They claim this new profile change was associated with a small, $\sim 0.9$\,nHz glitch in the pulsar spin frequency and the post-event evolution of $\dot{\nu}$ is consistent with the presence of creeping superfluid vortices at the crust-core boundary \citep{Alpar1984a, Alpar1984b, Akbal2017}.

This work builds upon the previous analyses of the profile and spin-down variability of PSR~J0748$-$4042 by \citet{Brook2014, Brook2016}.
We present an even more extensive dataset that covers a 50-year timespan using the four different radio observatories listed in Section~\ref{sec:obs}.
Our framework for detecting and analysing changes in the pulse profile shape and rotational properties of the pulsar are outlined in Section~\ref{sec:method}.
In Section~\ref{sec:poln} we describe the polarisation properties of the pulsar.
We present updated timing measurements in Section~\ref{sec:timing} and revisit the claimed glitch association with the 2015 profile/spin-down event.
Section~\ref{sec:variability} describes the results of running the Gaussian process regression technique on the collected timing and profile residuals to study the rotational and emission properties of the pulsar over time. 
This includes the first extension of the profile analysis to the polarisation domain.
We discuss the implications of our analysis on the correlated spin-down and profile shape changes detected in PSR~J0738$-$4042 in Section~\ref{sec:discussion} and prospects for future work in Section~$\ref{sec:conclusion}$.

\section{Observations}\label{sec:obs}

PSR J0738$-$4042 has been extensively monitored since its initial discovery.
To develop a complete understanding of the pulsar's rotational history over this time period, we have combined the timing datasets from both historical and ongoing timing campaigns from 1972 to 2023.
Here we give an overview of the four radio observatories that contributed to this unique, 50-year long data set.

\subsection{Deep Space Network}

Regular timing observations of PSR J0738$-$4042 were carried out using four antennas of NASA's Deep Space Network (DSN) from 1972 to 1983 \citep{Downs1983, Downs1986}.
They were primarily carried out with the 26-m DSS-13 and 64-m DSS-14 antennas\footnote{The original DSS-13 antenna was replaced by a 34-m dish in 1991 and the primary reflector of DSS-14 was extended to 70-m in 1988.} located in Goldstone, USA.
Supplementary timing data were also collected with the 26-m DSS-11 (Goldstone, USA) and DSS-62 (Madrid, Spain) antennas.
All four antennas recorded left-hand circularly polarised radio waves  at a central frequency of 2388\,MHz.
The pulse times of arrival (ToAs) recorded by \citet{Downs1983} and \citet{Downs1986} were reported as geocentric arrival times after being corrected from UTC to a uniform `ephemeris time' (ET).
We used their listed ET to UTC corrections to revert the ToAs back to UTC before converting them into a format compatible with the {\sc tempo2} pulsar timing package \citep{Hobbs2006}.
\subsection{Hartebeesthoek}

PSR J0738$-$4042 was observed using the 26-m antenna at HartRAO (formerly known as DSS-51) from 1984 to 2012 every 1-14 days \citep{Brook2014}.
These observations were performed using the 13-cm and 18-cm receivers, where left-hand circularly polarised radio waves were recorded at central frequencies of 1644\,MHz and 1668\,MHz with the 18-cm receiver, while the 13-cm receiver operated at 2270\,MHz and 2273\,MHz.
The signal processor in use before 2003 April provided a single frequency channel covering 10\,MHz of bandwidth for both receivers. 
After this date, an upgraded signal processor was installed allowing multiple frequency channels to be recorded over 8\,MHz of bandwidth with the 18-cm receiver and 16\,MHz when using the 13-cm receiver.
More specific details on the pulsar recording system can be found in \citet{Flanagan1995}.

\subsection{Parkes}

\begin{figure}
    \centering
    \includegraphics[width=\linewidth]{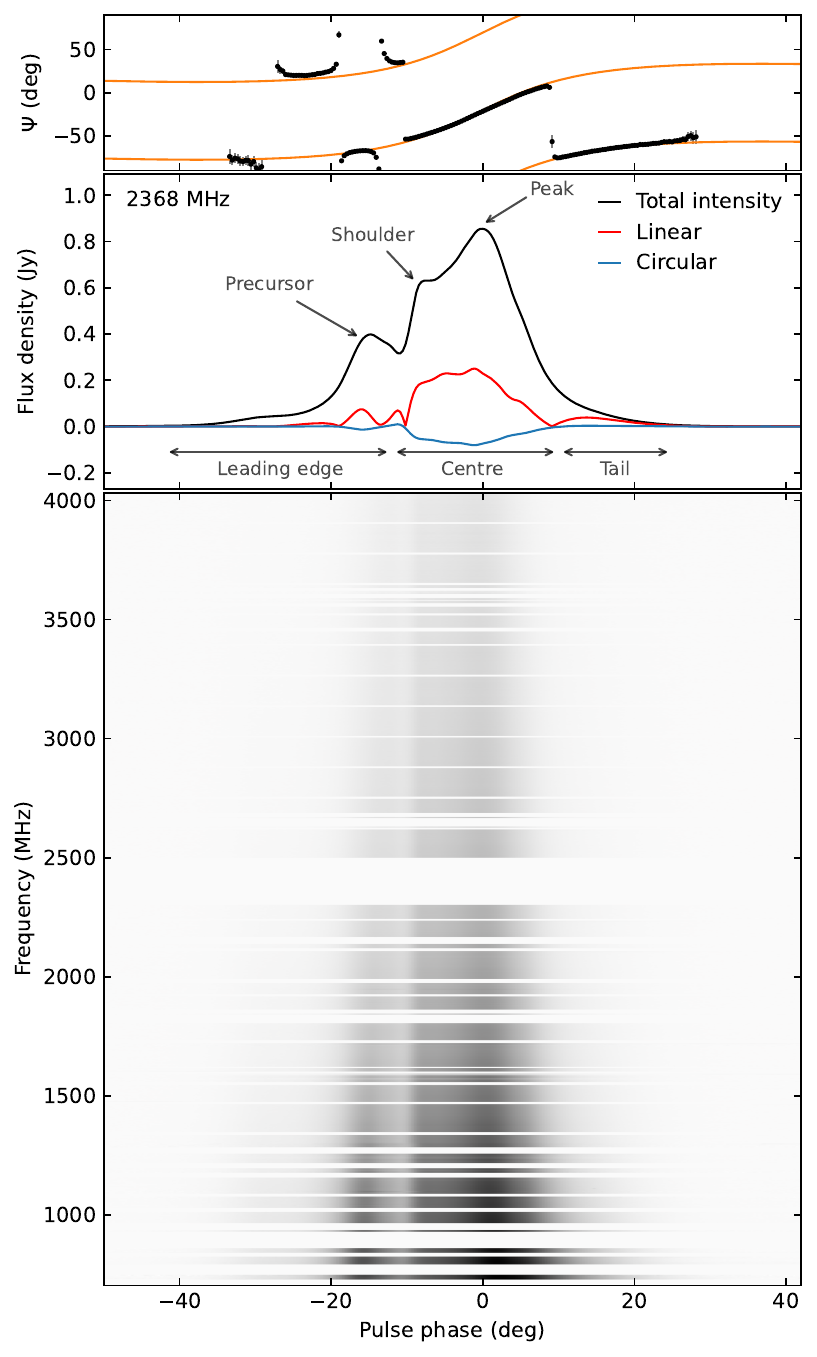}
    \caption{Time- and frequency-averaged polarisation pulse profile and total intensity spectrum of PSR~J0738$-$4042 observed with the Parkes UWL receiver. The upper panel shows the linear polarisation position angle with the best-fit rotating vector model overlaid in orange. The middle panel displays total intensity in black, linear polarisation in red and circular polarisation in blue, with profile components of interest highlighted by arrows. The lower panel shows the normalized total intensity as a function of pulse phase and frequency, where channels that were affected by radio interference are not plotted.}
    \label{fig:profile}
\end{figure}

Sparse observations of PSR J0738$-$4042 were performed using \textsl{Murriyang}, the 64-m radio telescope at the Parkes Observatory, since the early 1970s \citep{Komesaroff1970, Backer1976b, McCulloch1978, Manchester1980}.
Between 1997 and 2008, the pulsar was observed intermittently at 1369\,MHz using the central beam of the Parkes Multibeam receiver \citep{Staveley-Smith1996}, with dual linear polarisation data captured using the analogue filterbank signal processors that were employed at the time. 
Approximately monthly observations of PSR J0738$-$4042 have been taken since 2008 under the Parkes young pulsar timing programme in support of the \textsl{Fermi} mission  \citep{Weltevrede2010, Johnston2021}. 
Observations from 2008--2019 were primarily performed using the pulsar digital filterbank (PDFB) signal processors to capture data from the Multibeam receiver at a centre frequency of 1369\,MHz covering 256\,MHz of bandwidth, with a break in 2016 where the H-OH receiver (centre frequency of 1457\,MHz with 512\,MHz bandwidth; \citealt{Granet2011}) was used in place of the Multibeam receiver.
From 2019 onward, data were collected using the Ultra-Wideband Low (UWL) receiver which enabled recording of data across a continuous bandpass covering 704--4032\,MHz using the {\sc medusa} GPU-based signal processor \citep{Hobbs2020}.
A smaller 256\,MHz band at a central frequency of 1369\,MHz was recorded simultaneously with PDFB4 up to 2021.
The effects of Faraday rotation from the magnetized interstellar medium were corrected for by de-rotating the Stokes $Q$ and $U$ components by the catalogued rotation measure of $12.1$\,rad\,m$^{-2}$ \citep{Noutsos2008}.
We present an example polarisation pulse profile and total intensity spectrum in Figure~\ref{fig:profile} after averaging together 13\,hrs of Parkes UWL observations, highlighting various profile features of interest. 
From this profile we measured an updated, structure-optimised dispersion measure (DM) of $160.49 \pm 0.02$\,pc\,cm$^{-3}$ that we obtained by fitting the average total intensity phase-frequency profile using the {\sc dm\_phase} package \citep{Seymour2019}, which we subsequently applied to all of our Parkes observations.
After updating the DM, we split the wideband {\sc medusa} data into eight separate subbands that were subsequently averaged in frequency to form multi-band profiles.
All data are written as {\sc psrchive} format archive files \citep{Hotan2004, vanStraten2012}.

\subsection{Molonglo}

Observations of PSR J0738$-$4042 using the 1.6\,km long East-West arm of the upgraded Molonglo Observatory Synthesis Telescope (UTMOST-EW; \citealt{Bailes2017}) began in 2015 March and ranmuntil 2021 January, after which the observations transitioned over to a recently refurbished 100-m sub-section of the North-South arm (UTMOST-NS) which began taking observations in 2021 July.
The UTMOST-EW arm recorded right-hand circularly polarised radio waves, covering 31\,MHz of bandwidth at a central frequency of 835\,MHz. The UTMOST-NS system records dual linear polarisation data covering 45\,MHz of bandwidth centered on 831\,MHz \citep{Day2022}.
All data were coherently dedispersed \citep{Hankins1975} at the DM of the pulsar and saved in {\sc psrchive} format.
The data were cleaned using a bespoke alogorithm {\sc xprof} (Bailes, M., priv. comm.), which makes use of a nearly noise-free profile for each pulsar in the timing program. An ephemeris for the pulsar and this profile allow cleaning of the archive data in sub-integrations versus phase and in frequency channels versus phase, predominantly in off-pulse regions for each observation. 
Further details regarding cleaning and processing of these data can be found in \citet{Jankowski2019} and \citet{Lower2020}.

\section{Methods}\label{sec:method}

\subsection{Pulsar geometry}

For a dipolar magnetic field topology where the radio emission is emitted along the field lines at low altitude, the sweep of the linear polarisation position angle ($\Psi$) is a purely geometric effect that traces the sky-projected direction of the pulsar magnetic field. This can be fit via the rotating vector model (RVM) of \citet{Radhakrishnan1969}
\begin{equation}
    \tan(\Psi - \Psi_{0}) =  \frac{\sin\alpha \sin(\phi - \phi_{0})}{\sin\zeta \cos\alpha - \cos\zeta \sin\alpha \cos(\phi - \phi_{0})},
\end{equation}
Here, $\alpha$ is the angle between the magnetic and rotational axes of the neutron star and $\zeta = \alpha + \beta$, where $\beta$ is the impact parameter between the magnetic pole and our line of sight. 
$\phi_{0}$ is the phase at which the inflection point of the PA swing ($\Psi_{0}$) occurs.

We use the method described in \citet{Johnston2023} to fit the RVM to the data. 
To summarise, the input observations are a set of linear polarisation position angles (PAs) and associated values of pulse phase, $\phi$. 
The PAs are derived from Stokes $Q$ and $U$, and are only determined when the linear polarisation was more than $3$-$\sigma$ above the off-pulse baseline. 
The peak of the  total intensity profile was (arbitrarily) chosen as the zero point of pulse phase. 
RVM fitting was carried out using a least-squares procedure implemented in {\sc Python}. 
A total of 85 evenly spaced trials in $\alpha$ were used ranging from 5\degr\ to 175\degr\ and in $\beta$, 40 trials were used between 0\degr\ and 20\degr. 
For each $\alpha$-$\beta$ pair, the {\sc scipy} routine {\sc least\_squares} was used to determine the values of $\phi_0$ and $\Psi_0$ via $\chi^2$ minimisation and the minimum $\chi^2$ was recorded. 
We note that PAs can only lie between $-90$\degr\ and $+90$\degr\ and hence the phase-wraps need to be taken into account.
In addition, radio emission can appear in two orthogonal polarisation modes (OPMs; \citealt{Manchester1975}), and these so-called orthogonal mode transitions must also be built into the fitting routine.

\subsection{Profile variability}

Following the method outlined in \citet{Brook2016}, we created profile variability maps from the Parkes 20-cm and UWL-multiband data that show the changes in profile intensity as a function of pulse phase and observing date.
We first removed any observations where the pulsar was not clearly detected or showed evidence of distortions due to instrumental artefacts or excess RFI that could not be excised.
Profiles with fewer than 1024 phase bins were also ignored.
An initial alignment was performed by calculating the correlation between an observation and a template as
\begin{equation}
    \mathcal{O} = \frac{\langle a | b \rangle}{\sqrt{\langle a | a \rangle \langle b | b \rangle}}.
\end{equation}
Here, $a$ is the observed profile we aligned against a reference profile, $b$, and $\langle a | b \rangle$ is the noise-weighted inner product given by
\begin{equation}
    \langle a | b \rangle = 4{\rm Re} \int_{0}^{\infty} \frac{\tilde{a}(f) \tilde{b}^{*}(f)}{S(f)} df,
\end{equation}
where $\tilde{a}(f)$ is the Fourier transform of profile $a$, $\tilde{b}^{*}(f)$ is the complex conjugate of profile $b$ in the Fourier domain and $S(f)$ is the power-spectral density of the noise.
For simplicity, we assume the noise is stationary and Gaussian with $S(f) \approx 1$.
The reference profile $b$ was chosen to be the observation with the highest signal-to-noise ratio (S/N). 

We then applied a second iteration of profile alignment based on the improved method of \citet{Brook2018}, where the profiles are aligned and normalized to sub-components that remain stable over time.
This approach maximises the number of pulse phase bins that are in agreement between an observation and the reference template (again, the highest-S/N observation) through a combination of phase shifts and an iterative scaling procedure.
The profiles were shifted in steps of one phase bin and multiplied by a scale factor that increases/decreases the profile peak by $\pm 5$\,per cent, after which the template was subtracted and the mean of the residuals computed.
Phase bins where the residuals are more than two standard deviations away from the mean are then removed from the profile and template, following which the residuals and their mean are re-calculated.
This process continues until the residual mean is $\lesssim 0.1$\,per cent of the previous value and the final number of remaining phase bins is recorded.
The optimal phase-shift and scale factor are then taken from the iteration that minimises the mean of the residuals normalized by the final number of remaining phase bins.

Once the profiles have been aligned, we then obtain our set of profile residuals by creating a `median' profile template that is comprised of the median normalized flux at each phase-bin, and then subtracting it from our observations.
We then created profile variability maps by using Gaussian process regression to fit non-parametric models to the profile residuals within an on-pulse window covering the $-45.7\degr$ to $29.5\degr$ pulse phase range in Figure~\ref{fig:profile}.
The Gaussian process regression was performed using an exponential-squared kernel that was computed via the {\sc george} library \citep{Ambikasaran2015}, where a $\chi^{2}$ minimisation approach was used to determine the model hyper-parameters.

\subsection{Pulsar timing and spin-down variations}

\begin{figure}
    \centering
    \includegraphics[width=\linewidth]{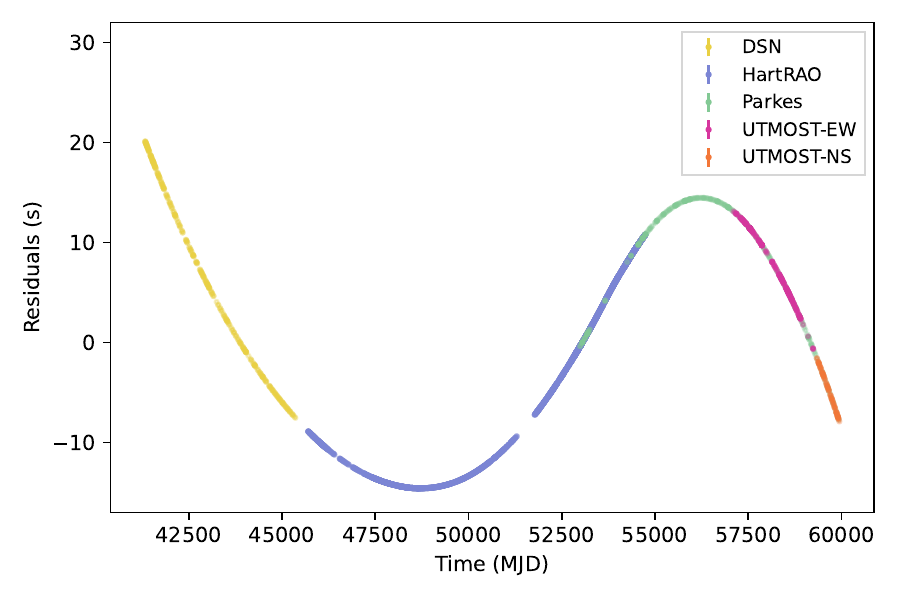}
    \caption{Timing residuals of PSR~J0738$-$4042 after fitting for the spin-frequency and spin-down rate of the pulsar. Each colour corresponds to observations taken with a different telescope.}
    \label{fig:timing}
\end{figure}

In order to search for spin-down variations that coincide with identified changes in the pulse profile, we required an up-to-date, phase-coherent timing solution across the full 50 years of PSR~J0738$-$4042 monitoring.
We used the {\sc tempo2} \citep{Hobbs2006, Edwards2006} pulsar timing software to assign integer pulse numbering to each ToA, which can be used as a reference when fitting a timing model to observations where there is an ambiguous integer number of rotations between them.
Jumps were included to account for differences in the standard templates used for obtaining ToAs and systematic offsets between telescope systems.

In Figure~\ref{fig:timing} we show the timing residuals of PSR~J0738$-$4042 after subtracting a timing model where the rotational evolution of the pulsar is described by an initial spin-frequency and spin-down rate referenced to MJD 51700.
The residuals are dominated by a structure that resembles a cubic polynomial, which could be interpreted as a significant second spin-frequency derivative ($\ddot{\nu}$) being present in the timing.
Measurements of the pulsar spin frequency and its first two deviates can be used to measure the observed braking index as
\begin{equation}\label{eqn:brake}
    n = \frac{\nu\,\ddot{\nu}}{\dot{\nu}^{2}},
\end{equation}
which can be used to infer the mechanism driving the braking of the pulsar.
However, similar cubic structures are ubiquitous in long-term timing datasets, resulting from the presence of strong timing noise \citep{Hobbs2010} as well as recoveries from large pulsar glitches that occurred before the first timing observations \citep{Johnston1999, Lower2021b}. 
Timing noise can also introduce significant biases in our timing models \citep{vanHaasteren2013}.
We avoided these issues by using the {\sc TempoNest} Bayesian pulsar timing plugin \citep{Lentati2014} to {\sc tempo2} which uses the {\sc MultiNest} nested sampling algorithm \citep{Feroz2009} to simultaneously fit the sky position, proper motion, $\nu$, $\dot{\nu}$, $\ddot{\nu}$ and timing noise properties of PSR~J0738$-$4042.
The timing noise is modelled in the Fourier domain as a red power-law process, the power spectral density of which is given by
\begin{equation}
    P(f) = \frac{A_{\rm r}^{2}}{12\pi^{2}} \Big( \frac{f}{1\,{\rm yr}} \Big)^{-\beta_{\rm r}}\,{\rm yr}^{3},
\end{equation}
where $A_{\rm r}$ and $\beta_{\rm r}$ are the amplitude and spectral index of the noise respectively.

After obtaining an updated timing model, we then used the non-parametric modelling technique developed by \citet{Brook2014, Brook2016, Brook2018}, which uses Gaussian process regression with a squared exponential kernel to fit the timing residuals.
A key assumption that underpins this method is that all unmodelled changes in the timing of a pulsar correspond to variations in the spin-down rate, where $\dot{\nu}(t)$ is simply the second derivative of the covariance kernel.
As with the profile variability maps, we used {\sc george} to generate the non-parametric models.
We sampled the model hyper-parameters by using the Bayesian Inference Library, {\sc bilby} \citep{Ashton2019}, as a wrapper for the {\sc PyMultiNest} nested sampling algorithm \citep{Feroz2009, Buchner2014}.
After fitting the residuals, we generated $\dot{\nu}$ points at the epoch of each timing observation.
Our uncertainties on $\dot{\nu}$ were obtained by taking 1000 random draws from the hyper-parameter posteriors and then taking the 2.5 and 97.5\,percentiles of the corresponding distribution of spin-down values at each epoch.

\section{Polarisation properties}\label{sec:poln}
The polarisation properties of PSR~J0738--4042 across the UWL band have been discussed recently by \citet{Sobey2021} and \citet{Oswald2023}. Observations at 8~GHz were shown in \citet{Johnston2006}. 
Figure~\ref{fig:profile} shows that the profile displays a low level of left-hand (negative) circular polarisation throughout and a moderate degree of linear polarisation, typical of middle-aged pulsars. 
There are 5 orthogonal mode jumps in the position angle sweep, described in \citet{Karastergiou2011}. 
\citet{Ilie2019} pointed out that apparent RM variations as a function of pulse phase were largest at the phases corresponding to the precursor.

\begin{figure}
    \centering
    \includegraphics[width=8.5cm]{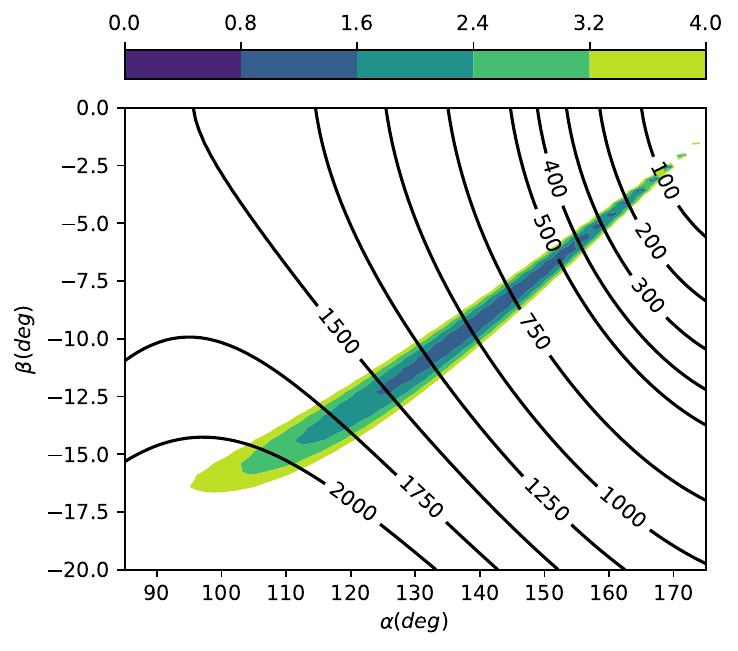}
    \caption{Results of RVM fitting to the PA swing of PSR~J0738--4042. The figure shows the $\alpha-\beta$ plane with the filled-contours denoting the $\chi^2$ of the fit. The black line-contours denote lines of constant emission height in km assuming a filled beam.}
    \label{fig:rvmfit}
\end{figure}

The best-fit RVM curve is displayed in the top panel of Figure~\ref{fig:profile}. Figure~\ref{fig:rvmfit} shows the $\chi^2$ contours in $\alpha-\beta$ space from the RVM fits. The results show that $\alpha$ must be greater than 100\degr\ and that $|\beta|$ cannot be larger than 17\degr. The inflection point is well determined; $\phi_0=-0.4\degr$, and at this location $\Psi_0 = -22\degr$. \citet{Blaskiewicz1991} showed that aberration and retardation effects cause the inflection point to lead the midpoint of the profile by an amount $\Delta_\phi$ given by
\begin{equation}
\label{eqn:bcw}
\Delta_\phi = \frac{8 \pi h_{\rm em}}{P c}
\end{equation}
with $h_{\rm em}$ the emission height, $P$ the spin period and $c$ the speed of light. In our case $\Delta_\phi=4.3\degr$ and therefore $h_{\rm em} = 330$~km. Coupled with the width of the pulse profile, this value of emission height traces out a contour in $\alpha-\beta$ space (see Figure~\ref{fig:rvmfit}) and this contour intersects the $\chi^2$ of the fitting at $\alpha=160\degr$, $\beta=-7\degr$.
The half-opening angle ($\rho$) of the radio beam is given by
\begin{equation}\label{open_height}
    \rho = 3 \sqrt{\frac{\pi h_{\rm em}}{2P\,c}},
\end{equation}
which yields $\rho=12\degr$, meaning that the line-of-sight cuts about half-way down the beam.

\section{Updated pulsar timing}\label{sec:timing}

We used {\sc TempoNest} to sample the astrometric (position and proper-motion), rotational ($\nu$, $\dot{\nu}$ and $\ddot{\nu}$), and stochastic timing noise properties of PSR~J0738$-$4042.
In order to speed up the parameter estimation, we only included the DSN, 13-cm HartRAO and 20-cm Parkes ToAs.
An additional 0.1\,ms uncertainty was added in quadrature to the DSN and Parkes ToAs to account for excess scatter originating from a combination of underestimating the formal ToA uncertainties and low-level, stochastic profile shape variations.

We tested whether the data favours a $\ddot{\nu}$ term by comparing the evidences from two separate {\sc TempoNest} fits: one where $\ddot{\nu}$ was a free parameter and another where it was fixed at zero. 
The resulting log Bayes factor of $\ln\mathcal{B} = 66$ indicates there is overwhelming evidence for a non-zero $\ddot{\nu}$ in our timing of PSR~J0738$-$4042 \citep{Skilling2004}.
Using the inferred values of $\nu$, $\dot{\nu}$ and $\ddot{\nu}$, we obtain from Equation~\ref{eqn:brake} a braking index of $n = 23300 \pm 1800$.
This value is much larger than the typical braking indices observed in young pulsars \citep{Espinoza2017, Parthasarathy2020, Lower2021b}. 
Our apparent braking index is a factor of four times smaller and of the opposite sign to that reported in \citet{Lower2020}, but within a factor of two of the pre-2015 event value obtained by \citet{Zhou2023}.
To assess the stability of the apparent $\ddot{\nu}$ over time, we also performed two additional {\sc TempoNest} fits to segments of the data on either side of the large profile/timing event reported by \citet{Brook2014}.
Our fits to the timing between 1972--2003 and 2009--2023 resulted in values of $n = 1500 \pm 600$ and $n = -6000 \pm 3000$ respectively. 

Non-stationary timing noise, which these spin-down events form part of, can potentially bias the pulsar parameters that we recover \citep{Keith2023}.
Since the timing of PSR~J0738$-$4042 is largely stable either side of the 2005 spin-down event, we could bypass its impact by ignoring the ToAs obtained between MJD 53563--55010 and adding a step-change in the spin-down rate ($\Delta\dot{\nu}$).
Our resulting measurements of the pulsar properties are presented in Table~\ref{tbl:ephem}.
In this case, the recovered value of $\ddot{\nu} = (-4 \pm 6) \times 10^{-26}$\,s$^{-3}$ is consistent with zero.
This confirms that the previously identified value of $n = 23300 \pm 1800$ is due to the $\dot{\nu}$ event, while the positive and negative pre-/post-event braking indices effectively cancel each other out.
The change in spin-down rate of $\Delta\dot{\nu} = (1.70 \pm 0.09) \times 10^{-15}$\,s$^{-2}$ matches the overall change reported by \citet{Brook2014}.

\begin{table}
\begin{center}
\caption{Timing ephemeris of PSR~J0738$-$4042. Values in parenthesis indicate the 1-$\sigma$ uncertainties on the last digit. \label{tbl:ephem}}
\renewcommand{\arraystretch}{1.2}
\setlength{\tabcolsep}{4pt}
\begin{tabular}{ll}
\hline
Parameter (units) & Value \\
\hline
R. A. (J2000)                           & $07$:$38$:$32.317(1)$ \\
Decl. (J2000)                           & $-40$:$42$:$40.19(1)$ \\
$\nu$ (Hz)                              & $2.667235748(5)$ \\
$\dot{\nu}$ (s$^{-2}$)                  & $-1.150(3) \times 10^{-14}$ \\
$\ddot{\nu}$ (s$^{-3}$)                 & $-4(6) \times 10^{-26}$ \\
$\mu_{\alpha}$ (mas yr$^{-1}$)          & $-37.0(9)$ \\
$\mu_{\delta}$ (mas yr$^{-1}$)          & $30(1)$ \\
Timing epoch (MJD TDB)                  & 51700 \\
Position epoch (MJD TDB)                & 51360 \\
DM (pc\,cm$^{-3}$)                      & $160.49(2)$ \\
Time span (MJD)                         & 41331--59930 \\
Number of ToAs                          & 2635 \\
Solar system ephemeris                  & DE436 \\
$\log_{10}(A_{\rm r})$ (yr$^{3/2}$)     & $-10.17(4)$ \\
$\beta_{\rm r}$                         & $5.5(1)$ \\
\hline
\end{tabular}
\renewcommand{\arraystretch}{}
\end{center}
\end{table}





\subsection{Pulsar astrometry and emission geometry}

Our inferred proper-motion deviates significantly from that of \citet{Brisken2003}, who used imaging observations from the Karl G. Jansky Very Large Array (VLA) to obtain values of $\mu_{\alpha}= -14 \pm 1$\,mas\,yr$^{-1}$ and $\mu_{\delta} = 12 \pm 2$\,mas\,yr$^{-1}$.
Instead, it is about mid-way between the VLA measurement and previously reported pulsar timing values of $\mu_{\alpha} = -56 \pm 8$\,mas\,yr$^{-1}$ and $\mu_{\delta} = 46 \pm 9$\,mas\,yr$^{-1}$ obtained by \citet{Downs1983}.
Imaging at the low declination of PSR~J0738$-$4042 with the VLA is challenging, and the discrepancy between our measurement and that of \citet{Brisken2003} may be the result of underestimated uncertainties in the pulsar position in the radio images. 
Although we are able to somewhat bypass the effects of the 2005 spin-down event, there is a possibility that our timing uncertainties are under-estimated which can arise if the remaining timing noise is not perfectly modelled by a red power-law spectrum.
We further tested this by breaking the dataset up into four $\sim$10-yr blocks of timing spanning 1972--1982, 1982--1992, 1992--2002 and 2012--2023.
The ToAs within these blocks were then fit for the pulsar position and rotational parameters in the absence of a proper motion.
A non-linear least-squares fit to the recovered pulsar positions returned proper motion values of $\mu_{\alpha} = -37 \pm 1$\,mas\,yr$^{-1}$ and $\mu_{\delta} = 30 \pm 1$\,mas\,yr$^{-1}$.
This is consistent with the proper motion we infer from the full 50-yr dataset.
We also performed separate proper-motion fits using each $\sim$decade long set of ToAs.
The recovered posterior distributions had median values that differed from those listed in Table~\ref{tbl:ephem} by up to $+21/-17$\,mas\,yr$^{-1}$ in $\mu_{\alpha}$ and $+2/-22$\,mas\,yr$^{-1}$ in $\mu_{\delta}$, however with uncertainties that are consistent with the global fit at the 97.5\,per cent confidence interval.
Shifts in the centroids of the proper motion posteriors may be due to differences in the non-stationary timing noise of PSR~J0738$-$4042 sampled within each time span.

An additional limiting factor in studying the astrometric properties of PSR~J0738$-$4042 comes from the lack of an accurate distance measurement.
From the dispersion measure of 160.49\,pc\,cm$^{-3}$, the NE2001 \citep{Cordes2002} and YMW16 \citep{Yao2017} Galactic electron density models return distance estimates of 2.6\,kpc and 1.6\,kpc, respectively.
The larger NE2001 distance is consistent with the lower-limit of 2.1\,kpc inferred from \ion{H}{i} absorption along the line of sight to the pulsar \citep{Johnston1996}.
Assuming the NE2001 distance, our total proper-motion of $\mu_{\rm tot} = 47.7 \pm 0.9$\,mas\,yr$^{-1}$ corresponds to a two-dimensional transverse velocity of $588 \pm 11$\,km\,s$^{-1}$.
If this distance is correct, then PSR~J0738$-$4042 resides within the high-velocity tail of the pulsar population \citep{Hobbs2005}.

Assuming PSR~J0738$-$4042 was born in the Galactic mid-plane, then the time taken for the pulsar to reach its current position below the Galactic plane ($z \sim -400$\,pc), and therefore its approximate kinematic age, can be obtained by dividing the pulsar proper-motion by its position in Galactic latitude.
For PSR~J0738$-$4042, the position and proper-motion of the pulsar is $(l, b) = (254.2\degr, -9.2\degr)$ and $(\mu_{l}, \mu_{b}) = (-43.9\,{\rm mas\,yr^{-1}}, -18.6\,{\rm mas\,yr^{-1}})$.
This results in a kinematic age of $\tau_{\rm kin} \approx 1.8$\,Myr, which is less than half the `characteristic' age of $\tau_{c} = \nu/2|\dot{\nu}| = 4.3$\,Myr.
Differences in kinematic versus characteristic age estimates are not unusual (see e.g. \citealt{Noutsos2013}) and predominantly reflect the simplifying assumptions that go into computing characteristic ages, i.e. high initial spin-frequencies and rotational evolution purely due to magnetic dipole braking.

Previous works have shown there is a strong correlation between pulsar spin orientation and their proper-motion direction for pulsars with ages less than approximately 1~Myr \citep{Johnston2005, Noutsos2013}. 
Assuming the inflection point of our RVM fit to the phase-resolved linear polarisation position angle of PSR~J0738$-$4042 corresponds to the location of the magnetic pole, then the absolute position angle of the linear polarisation is $-22$\degr.
By comparison, the polarisation position angle of the proper-motion with respect to North on the sky is $-50$\degr. 
We should not read too much into the $\sim28\degr$ offset between the two vectors; given the pulsar's large kinematic age, its current proper motion vector may not reflect that of its birth vector (see \citealt{Noutsos2013}).

\subsection{Limits on pulsar glitches}

\begin{figure*}
    \centering
    \includegraphics[width=\linewidth]{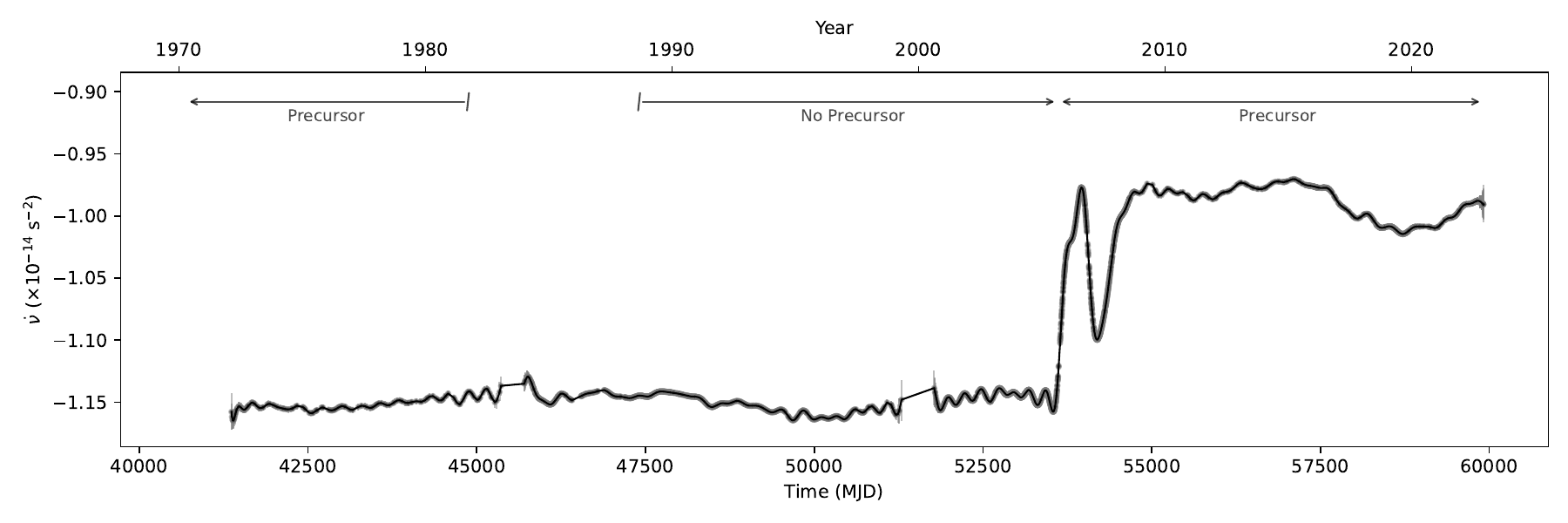}
    \caption{Spin-down evolution of PSR~J0738$-$4042 over 50 years of timing. Labels and time range indicated along the top correspond to when the precursor profile component in Figure~\ref{fig:profile} was or was not detected.}
    \label{fig:long_var}
\end{figure*}

A recent analysis of Parkes and UTMOST timing of PSR~J0738$-$4042 by \citet{Zhou2023} claimed that a small glitch took place on MJD $57359(5)$. 
The fractional changes in spin-frequency and spin-down rate of $\Delta\nu/\nu = 0.36(4) \times 10^{-9}$ and $\Delta\dot{\nu}/\dot{\nu} = -3(1) \times 10^{-3}$ would place this event among the smallest glitches recorded in a non-recycled pulsar to date \citep{Manchester2005, Espinoza2011, Lower2021b, Basu2022}. 
However, previous reports of similarly small glitches in other pulsars were often later found to be indistinguishable from fluctuations induced by timing noise following more detailed modelling (e.g., \citealt{Lower2020}).
Indeed, a search for glitches in UTMOST-monitored pulsars by \citet{Dunn2022} failed to identify any significant glitch events in PSR~J0738$-$4042.

We searched for evidence of this glitch by performing a Bayesian model selection study of the joint Parkes/UTMOST data covering the same MJD 57106--57878 range used in the \citet{Zhou2023} fit.
This includes Parkes H-OH observations that were not previously analysed in \citet{Zhou2023}.
As with our test for the presence of a $\ddot{\nu}$ in the longer timing dataset, we fitted the timing data with two separate models: one that included both $\Delta\nu$ and $\Delta\dot{\nu}$ glitch terms, and another where they are held fixed at zero.
Both fits included the red power-law timing noise model. 
We failed to recover a significant change in spin-frequency and instead set a 68\,per cent upper limit of $\Delta\nu/\nu \lesssim 2.6 \times 10^{-11}$, which is below the previously reported glitch amplitude. 
While our posterior distribution for the change in spin-down peaks away from zero with $\Delta\dot{\nu}/\dot{\nu} = 9(5) \times 10^{-3}$, our model comparison returned a log Bayes factor of $\ln\mathcal{B} = -20$ indicating the glitch model is heavily disfavoured when compared to the no-glitch model. 

The lack of any evidence for glitch activity in this pulsar over 50 years of monitoring is not surprising.
\citet{Fuentes2017} derived an approximate wait time between subsequent large ($\Delta\nu/\nu \sim 10^{-6}$) pulsar glitches of $\Delta T_{g} = 1/420\,{\rm Hz}^{-1} |\dot{\nu}|$. For PSR~J0738$-$4042, this corresponds to a wait time of approximately 7\,kyr.
While this relationship does not extend to small glitches, such as the one we have ruled out, \citet{Fuentes2017} found the rate of such events appears to track with increasing $\dot{\nu}$ despite challenges relating to sample completeness.

\section{Pulse profile and spin-down evolution}\label{sec:variability}

In Figure~\ref{fig:long_var} we present the recovered spin-down time series from fitting the updated timing residuals from all four observatories.
As in \citet{Brook2016}, we only plot spin-down values for dates where the pulsar was observed.
However, differences in observing systems, receiver availability and various upgrades make such a constraint impractical for this study.
There are two small gaps in our spin-down model between the end of the DSN observations and start of HartRAO monitoring between MJD 45360--45704 and an upgrade of the HartRAO system between MJD 51293--51777.
These gaps in coverage are sufficiently small ($\sim 1$\,yr) that we can confidently report that no large changes in the pulsar spin-down rate took place between the start of our timing dataset in January 1972 and the September 2005 event.
Instead, the pulsar displayed comparatively low-amplitude, low-fluctuation frequency $\dot{\nu}$ variations while remaining in the `high' pre-2005 spin-down state.
The impacts of both of the September 2005 and December 2015 profile/spin-down events are clearly visible \citep{Brook2014, Zhou2023}, along with a shorter-timescale, periodic wobble that may be due to small error in the pulsar position.

\begin{figure}
    \centering
    \includegraphics[width=\linewidth]{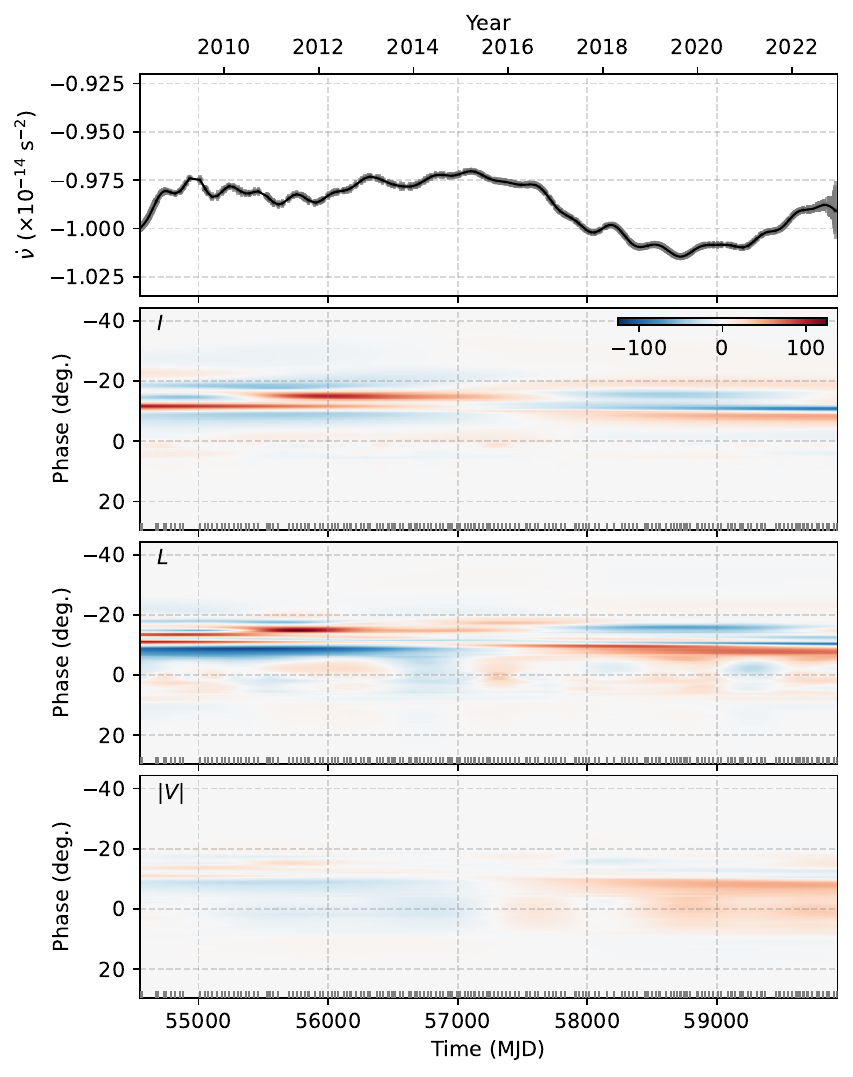}
    \caption{Spin-down rate (top) and Parkes 20-cm polarisation profile variation maps (middle to bottom) for total intensity, linear and absolute circular polarisation. The short ticks on the bottom-hand side of each variability map indicate the observation epochs.}
    \label{fig:pol_var}
\end{figure}

\begin{figure}
    \centering
    \includegraphics[width=\linewidth]{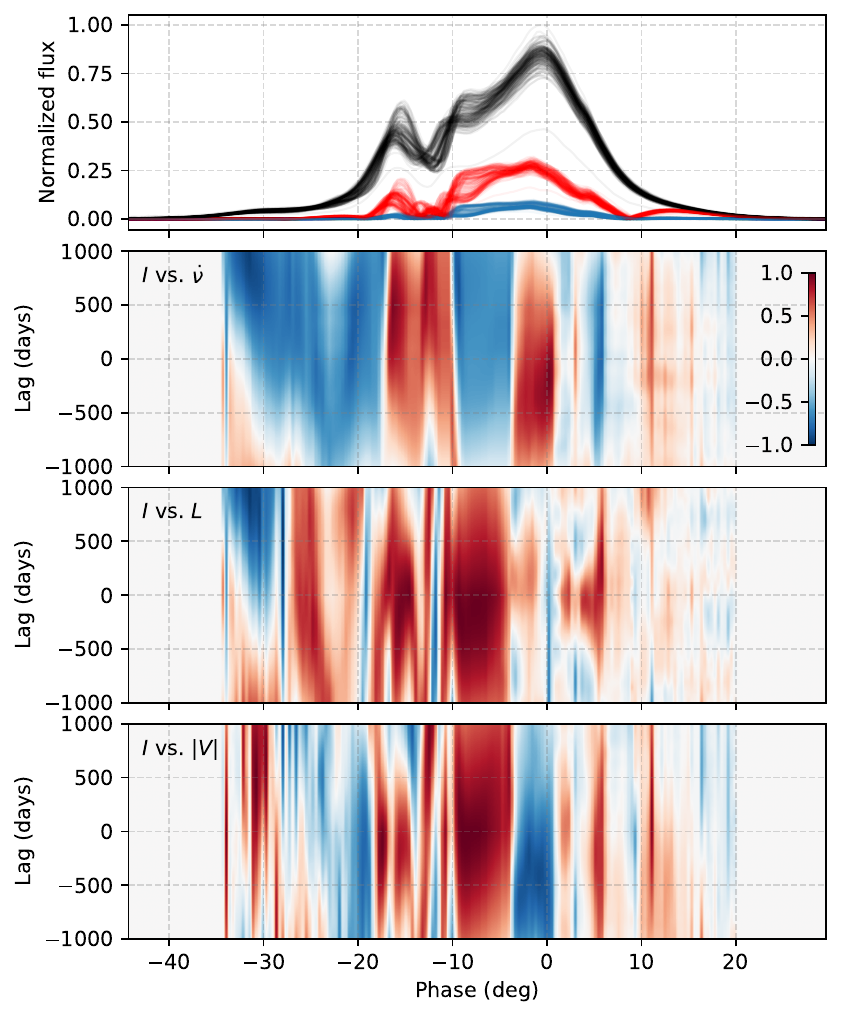}
    \caption{Parkes 20-cm polarisation profiles, shape/spin-down and polarisation correlation maps. Top panel shows the ensemble normalized total intensity (black), linear (red) and absolute circular (blue) polarisation profiles. Upper-middle panel shows the Spearman-Rank correlation coefficient between the profile shape and spin-down rate as a function of pulse phase and time-lag. The lower two panels show the Spearman-Rank coefficient from comparing the total intensity shape against the linear (lower-middle) and circular (bottom) profile shapes.}
    \label{fig:pol_corr}
\end{figure}

A comparison of the spin-down model against profile variability maps generated from the 2008--2023 Parkes 20-cm polarisation data is shown in Figure~\ref{fig:pol_var}.
The December-2015 event detected by \citet{Zhou2023} shows up as an increase in spin-down rate that coincides with a reduction in the precursor intensity and enhanced emission from the shoulder component.
Note that our 20-cm observations do not all share the same central frequency and bandwidth due to a combination of receiver/signal-processor availability and the upgrade to the UWL system.
Hence, some of the small scale variations during between 2016--2017 (MJD 57388--57754) and from 2021 (MJD 59215) onward may be impacted by frequency-dependent profile evolution \citep{Shaw2022}. 
The linear ($L$) and absolute circular ($|V|$) polarisation maps were generated using the phase shifts and profile scale factors obtained from aligning from the total intensity data. 
As in \citet{Brook2016}, we computed the Spearman rank correlation coefficient between the total intensity profile residuals and $\dot{\nu}$ for time-lags ranging between $\pm 1000$\,d.
We also performed a similar correlation analysis between total intensity with the absolute linear and circular polarisation residuals.
The resulting correlation maps are shown in Figure~\ref{fig:pol_corr} alongside the normalized polarisation profiles.

We confirm the December 2015 event found by \citet{Zhou2023} coincided with a reduction in relative precursor intensity and increase in shoulder component emission.
Compared to the 2005 event, the change in profile shape appears to have been relatively short-lived and is currently in the process of recovering back towards the pre-2015 shape.
This recovery coincides with a gradual reduction in |$\dot{\nu}$| from a peak value of $-1.01141(1) \times 10^{-14}$\,s$^{-2}$ on MJD 58724.
Similar to \citet{Brook2016}, we find the Stokes $I$--$\dot{\nu}$ correlation map in Figure~\ref{fig:pol_corr} shows a complicated relationship, with strong, alternating positive and negative correlations recovered across much of the profile.
Additionally, the lack of any strong periodic correlation at positive and negative lags adds further credence to the short-timescale, periodic signal in the $\dot{\nu}$ model being the result of an unaccounted error in the pulsar position.
The broader, more intense correlations than those reported in \citet{Brook2016} may be reflective of the longer timescale over which the \citet{Zhou2023} event takes place.
Overall, the changes in the linear polarisation profile largely followed the variations in total intensity.
Visually, there are small-scale circular polarisation variations, particularly at pulse phases corresponding to the precursor and shoulder components, that primarily track the changes in linear polarisation. In addition, there is a broader change from a circular deficit to an excess towards the middle of the variability map. This change in the sign of the variation indicates a long-term, secular change in circular polarisation; it appears in the middle of the data-span due to the subtraction of the median profile.
There are also short-term linear polarisation changes around the profile peak that are not reflected in either Stokes $I$ or $|V|$, which could be due to small changes in rotation measure along the line of sight through the interstellar medium. 
Indeed both the DM and RM of the pulsar have respectively decreased by approximately $0.2$\,pc\,cm$^{-3}$ and $2$\,rad\,m$^{-2}$ over the last $\sim 16$\,yr (Sobey et al. in prep).
The total intensity with linear and circular correlation maps in Figure~\ref{fig:pol_corr} display a substantial degree of commonality with one another.
As with the $\dot{\nu}$ map, the strongest correlations (both positive and negative) occur at zero-lag, indicating there is little to no lag between a change in the total intensity profile and the linear or circular polarisation components.

\subsection{Wideband profile variations}

\begin{figure*}
    \centering
    \includegraphics[width=\linewidth]{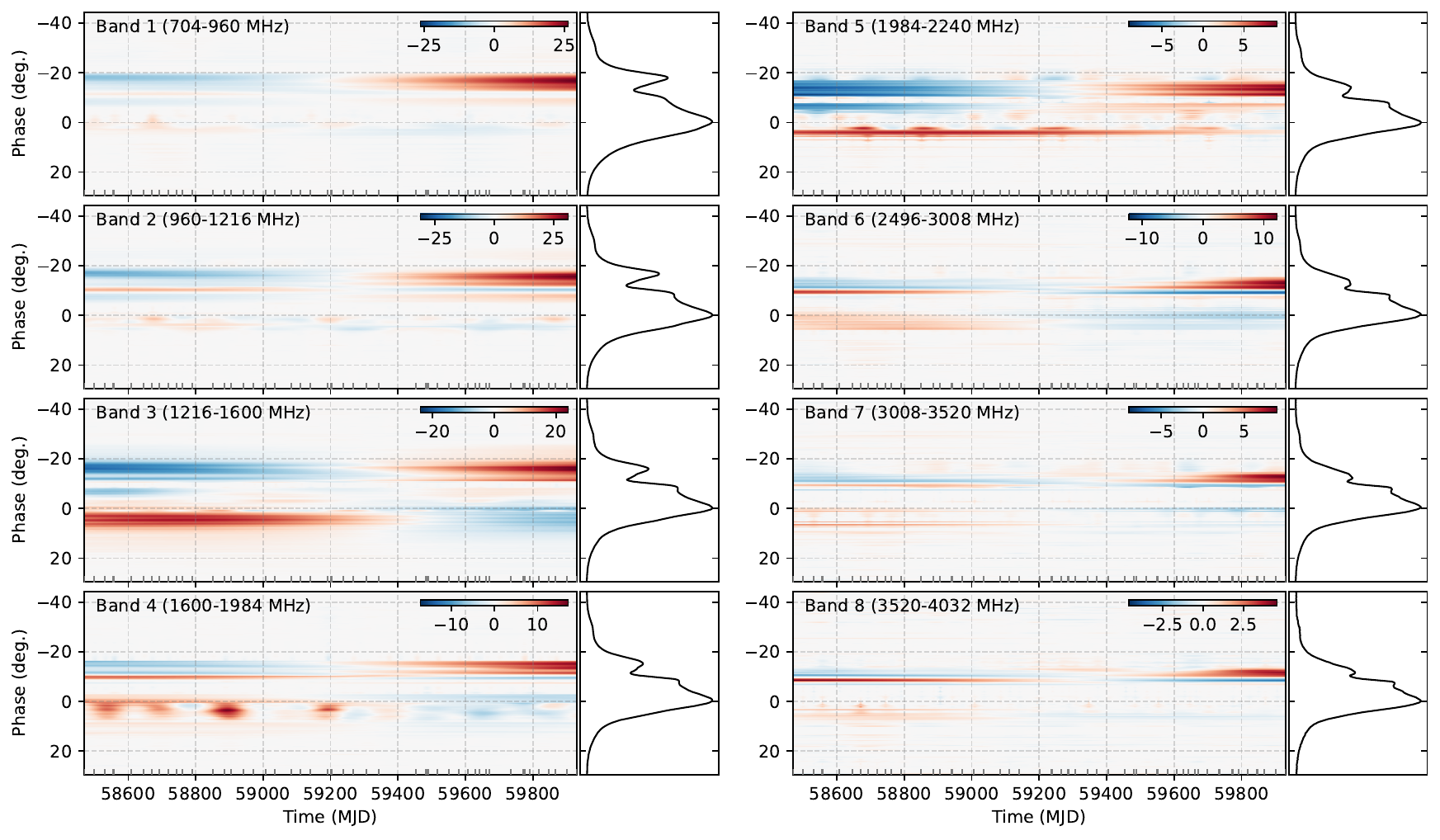}
    \caption{Total intensity variations across four years of Parkes-UWL observations split into eight subbands. Observation dates are indicated by gray vertical ticks. Note the ranges covered by each colorbar are subband specific.}
    \label{fig:uwl_var}
\end{figure*}

We further extended the profile variability method to the four years of Parkes-UWL observations spanning 2019--2023.
Unlike the polarisation analysis, we performed the phase alignment and scaling on each subband independently. 
The resulting multi-band profile variability maps are shown in Figure~\ref{fig:uwl_var} along with the corresponding median profiles.
Visually, the different variability maps all appear to display similar features. 
In particular, an initial deficit in the precursor emission that transitions into an excess around MJD 59300 appears in all eight subbands.
A long-term change in the peak with the opposite sense (initial excess turning into a deficit) also takes place along the same timescale, appearing most intense in Band 3 (1216--1600\,MHz). 

\begin{figure}
    \centering
    \includegraphics[width=0.7\linewidth]{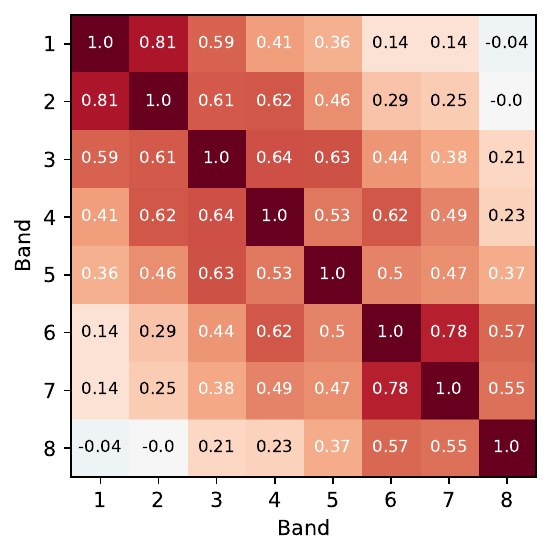}
    \caption{Heatmap of Pearson linear correlation coefficient between the UWL subband variability maps in Figure~\ref{fig:uwl_var}.}
    \label{fig:uwl_corr}
\end{figure}

The profile variations appear less intense at higher observing frequencies, though it is clear from the median templates that the relative intensities of the three dominant components vary substantially with frequency.
In particular, the precursor appears noticeably weaker when compared to the peak and shoulder at higher frequencies. 
This is also readily visible in Figure~\ref{fig:profile}.
Simple power-law fits to the average flux densities within windows placed around the precursor ($-20$\degr\ to $-11$\degr), shoulder ($-11$\degr\ to $-4$\degr) and peak ($-4$\degr\ to $6$\degr), where we obtained respective spectral indices of $-1.81 \pm 0.07$, $-1.40 \pm 0.05$ and $-1.55 \pm 0.05$.
The shallower spectral index of the shoulder accounts for it becoming the dominant component detected at 8\,GHz \citep{Johnston2006}.

We quantified the level of similarity between the polarisation variability maps by computing the Pearson linear correlation coefficient
\begin{equation}
    r_{\rm p} = \frac{N(\sum_{i}^{N} a_{i}\,b_{i}) - (\sum_{i}^{N} a_{i})(\sum^{N}_{i} b)}{\sqrt{N\sum_{i}^{N} a_{i}^{2} - (\sum_{i}^{N} a_{i})^{2}} \sqrt{N\sum_{i}^{N} b_{i}^{2} - (\sum_{i}^{N} b_{i})^{2}}},
\end{equation}
where $a$ and $b$ are the two variability maps that have been flattened from 2-D to 1-D arrays, and $N$ is the sample size.
Our resulting correlation coefficients between the eight subbands are presented in Figure~\ref{fig:uwl_corr}. 
The level of correlation between subbands drops significantly with increasing separation in frequency.
This is particularly obvious with Band 8 which becomes de-correlated when compared to Bands 1 and 2.
The mechanism behind this decreasing correlation is illustrated both in the lower panel of Figure~\ref{fig:profile} and the median templates shown in Figure~\ref{fig:uwl_var}, with the increase in separation between the precursor and shoulder components with decreasing observing frequency and aforementioned differences in component spectral indices.
Since most of the variability occurs in the precursor, its movement to later pulse phases reduces the level of correlation between bands and explains why profiles that are near in frequency have stronger correlations.
This behaviour is consistent with the expected narrowing of the radio beam with increasing frequency due to radius-to-frequency mapping \citep{Thorsett1991}.
Note that the lower sensitivity of the UWL near the top of the band may also account for some of the apparent decrease in variability.

\section{Discussion}\label{sec:discussion}

Observational phenomena such as emission state switching, nulling and pulse jitter demonstrate that the magnetosphere of a pulsar is incredibly dynamic, with turbulent plasma flows and both short- and long-timescale variations in the local magnetic field structure.
Observations of pulse microstructure demonstrate that variations in pulsar radio emission can occur on millisecond to nanosecond timescales \citep{Hankins1978}.
The ensemble effect of these short time-scale emission variations may contribute to the longer-timescale timing noise detected in many pulsars. 
Mode-changing seems ubiquitous across the pulsar population and is observed at both radio and X-ray wavelengths \citep{Hermsen2013, Hermsen2017, Hermsen2018}. 
The timescales for mode-changing range from only a few rotations of the star to many years in the intermittent pulsars, but what sets the timescale is unclear.
While the long timescale variability of mode-changing pulsars can be linked to short-term changes in their emission \citep{Stairs2019, Brook2019}, the profile and spin-down changes observed in PSR~J0738$-$4042 are somewhat different as they are both transient in nature and take place over timescales lasting many years to decades. 
Although this behaviour is likely magnetospheric in origin, the exact process that triggers it is unknown. 
A purely magnetospheric-driven trigger that results in an unusual form of mode-changing behaviour in which the profile shape changes are not always accompanied by a change in spin-down rate cannot be ruled out.  
In the following sub-sections we explore several potential explanations for driving the variable long-term rotational and emission properties of this pulsar.


\subsection{Origin of the second spin-frequency derivative}

Our timing fits to both the full 50-year dataset as well as smaller 30- and 14-year sections before and after the 2005 event returned extraordinarily large and disparate values of $n =23300 \pm 1800 $, $1500 \pm 600$ and $-6000 \pm 3000$ respectively.
Braking indices with large values and variable signs were previously attributed to model mis-specification, where the value of $\ddot{\nu}$ is biased by unmodelled timing noise \citep{Hobbs2010}.
Long-term glitch recoveries could also induce similarly anomalous braking indices \citep{Johnston1999}. 
Works by \citet{Parthasarathy2019, Parthasarathy2020} and \citet{Lower2021b} overcame the mis-specification issue by using a Bayesian pulsar timing and model selection framework. 
This led to the discovery of a sample of young pulsars with robust braking index measurements between $10 \lesssim n \lesssim 100$.
\citet{Parthasarathy2020} suggested these braking indices could result from a gradual alignment of the pulsar magnetic and spin axes over time (see \citealt{Johnston2017} and references therein). 
\citet{Lower2021b} demonstrated large values of $n$ in glitching and some (seemingly) non-glitching pulsars were consistent with arising from a linear spin-down recovery induced by superfluid vortex creep at the crust-core boundary \citep{Akbal2017}, where the inter-glitch evolution in spin-down is `reset' during a large glitch.
Despite employing the same model selection criterion, our value of $n$ from fitting the full timing of PSR~J0738$-$4042 is an extreme outlier, and reflects the impact that the large transient spin-down events have on its timing.
While the pre- and post-2005 event braking indices are less extreme, they are still significantly larger than the apparent braking indices of the pulsars studied by \citet{Parthasarathy2020} and \citet{Lower2021b}.

\citet{Parthasarathy2019} speculated an outlier in their pulsar sample (PSR~J0857$-$4424; $n = 2890$) could be due to the presence of a companion object with a wide orbital separation of tens of AU. 
We examine this possibility for PSR~J0738$-$4042. 
A search through the \textsl{Gaia} archive\footnote{\href{https://gea.esac.esa.int/archive/}{https://gea.esac.esa.int/archive/}} for catalogued sources finds that the closest star is Gaia DR3 5536901795956300160 (magnitude and colour $G = 20.84$ and $BP-RP = 1.11$ respectively) at a $\sim 7.6$\," offset from the position of the pulsar.
Assuming a distance of 2.6\,kpc, the projected separation would be $\sim$5.5\,pc, effectively ruling out any association between the two objects. 
Ignoring the effects of extinction and assuming the standard mass-luminosity relation of $L/L_{\odot} \approx 0.23 (M/M_{\odot})^{2.3}$ for zero-age main sequence stars with $M < 0.43 M_{\odot}$ \citep{Carroll2006}, we use the limiting G-passband magnitude of 22 for \textsl{Gaia} to compute a rough upper-limit of $0.24$\,$M_{\odot}$ on a possible stellar companion to PSR~J0738$-$4042. 
For a $0.24$\,$M_{\odot}$ companion in a circular orbit, we can use the prescription of \citet{Kaplan2016} to show that the orbital period would be $\sim 1100$\,yr with a separation of $\sim 125$\,AU.
Given the time dependence of the braking index combined with the aforementioned spin-down variability, we disfavour a stellar binary origin.

A collection of small bodies orbiting the pulsar in a debris disk would induce a power-law red noise process in the residuals through the superposition of the reflex motion imparted on the neutron star \citep{Shannon2013, Jennings2020}. 
The variance of the noise is expected to increase with time up until the timing baseline exceeds the maximum orbital period of bodies within the debris disk, after which the noise power spectrum would flatten and turn over.
While this could result in different values of $\ddot{\nu}$ being recovered for different segments of the data, it is clear from Figure~\ref{fig:long_var} that the time dependence of $\ddot{\nu}$ in PSR~J0738$-$4042 is imparted by the large spin-down/profile events in 2005 and 2015.
These events also make it impossible to distinguish the expected power-spectrum turnover from non-stationary timing events.
Even prior to the 2005 event, the pulsar displayed low-level changes in $\dot{\nu}$ which may be related to the disappearance of the precursor profile component in the early 1980s followed by enhanced shoulder emission from 1992 onward \citep{Brook2014}. 
Hence a $\ddot{\nu}$ originating purely from low-level, correlated spin-down and profile variations cannot be discounted.
Thermal emission from a sufficiently dense asteroid belt or debris disk around a pulsar could be detected as an infrared counterpart (as for e.g. the magnetar 4U~0142$+$61; \citealt{Wang2006}).
Currently, there is no reported infrared source at the position of PSR~J0738$-$4042.
Given the limitations imposed on our timing, future deep infrared or sub-mm observations may provide the best constraints on the existence of a debris disk around the pulsar.

\subsection{Correlated profile and spin-down events}

We used the Gaussian process regression techniques developed by \citet{Brook2014, Brook2016, Brook2018} to study the spin-down evolution of PSR~J0738$-$4042 over half a century and changes in the profile over the last 16 years.
While our profile analysis did not cover the initial 16-years of timing, previously published observations by the DSN and Parkes clearly showed that the precursor profile component was active throughout the 1970s and the early 1980s \citep{Komesaroff1970, Backer1976b, McCulloch1978, Downs1979, Manchester1980, Manchester1983}, with the last reported detection being in an 843\,MHz observation taken by the Molonglo telescope in 1981 \citep{Biggs1986}.
The lack of available profiles between 1981--1988 means it is unknown whether the precursor emission underwent a gradual decay into quiescence or was suddenly quenched.
It is also not clear whether the precursor shared the exact same polarisation properties in these early observations.
While the polarisation profiles published by \citet{Komesaroff1970} and \citet{Manchester1980} do not show evidence of OPMs in the PA swings, the limited S/N and low pulse phase resolution may have resulted in the jumps being unresolved.
However, the linear polarisation profile in Figure 4 of \citet{Manchester1980} does display the same multiple dips associated with the OPMs in Figure~\ref{fig:profile}, suggesting the precursor may have been orthogonally polarised during this previous appearance.

Our spin-down measurements in Figure~\ref{fig:long_var} demonstrate the precursor disappearance (sometime between 1981--1988, see \citealt{Brook2014}) was not associated with a large deviation in the spin-down rate. 
Instead, the spin-down appeared to experience only low-level variations with a standard deviation of $7 \times 10^{-17}$\,s$^{-2}$ about a mean value of $-1.147 \times 10^{-14}$\,s$^{-2}$.
The presence of the same precursor component in these archival profiles suggests the region of the magnetosphere responsible for the precursor component is intermittently active.
However, the lack of a substantial change in spin-down associated with the precursor disappearance is somewhat inconsistent with expectations for mode-changing, as the reduction in particle outflow should have resulted in a corresponding change in the torque acting to slow the neutron star \citep{Kramer2006}.

\citet{Zhou2023} suggested the December 2015 event in PSR~J0738$-$4042 was caused by a coupling of the pulsar magnetosphere to crust-quake induced glitch in the pulsar.
In principle, crust quakes and the slow decay of magnetic fields inside a neutron star can result in prolonged variations in the rotation and profile of a pulsar.
However, the associated phenomena are often short-lived with no detected change in profile shape.
For instance, the delayed spin-up glitches detected in PSR~J0534$+$2200 have rise times that have lasted up to 2.5\,d (see \citealt{Shaw2018, Shaw2021} and references therein).
While we were able to confirm the 2015 correlated profile/spin-down event in PSR~J0738$-$4042, our timing measurements failed to recover evidence for the reported glitch.
Combined with several profile shape changes that did not result in an appreciable change in spin-down rate, the correlated behaviour in this pulsar is therefore more likely to be driven by an external, magnetospheric process rather than an internal mechanism.

The 2005 profile and spin-down event was interpreted by \citet{Brook2014} as evidence of a $\sim 5 \times 10^{14}$\,g asteroid having interacted with PSR~J0738$-$4042.
Asteroids and other minor bodies that orbit too close to a pulsar could be disrupted through irradiation, after which their ionised remains would interact with the pulsar magnetosphere resulting in changes in spin-down and an alteration of the emission mechanism \citep{Cordes2008}.
Given the high transverse velocity of the pulsar, these in-falling asteroids are unlikely to be the remains of a relic disk that survived the supernova that created PSR~J0738$-$4042.
Instead, they could have formed from a disk of material that fell back toward the neutron star soon after its formation \citep{Michel1988}, or matter that was stripped from a companion star if the natal kick on the pulsar was in the right direction \citep{Hirai2022}.
A key prediction of the asteroid model is the altered profile and spin-down state should decay back to the original, pre-interaction state once the excess charged particles are exhausted.
Our negative braking index measured from the 2009--2023 timing could indicate the pulsar is undergoing a slow, secular return to the pre-2005 spin-down state, although this precludes the continued recovery of the spin-down back toward the pre-2015 event value.
Even if no secular decay in the weakened spin-down state has occurred, the initial step in $\dot{\nu}$ associated with the events can be used to infer the change in charged particle density within the magnetosphere via \citep{Kramer2006}
\begin{equation}\label{eqn:charge}
    \Delta\rho = \frac{3 I \Delta\dot{\nu}}{R_{\rm pc}^{4} B_{0}}.
\end{equation}
Here, $I = (2/5)M R^{2}$ is the moment of inertia, $R_{\rm pc} = \sqrt{2\pi R^{3} \nu/c}$ is the polar cap radius and $B_{0} = 3.2 \times 10^{19} \sqrt{-\dot{\nu}/\nu^{3}}$ is the surface magnetic field strength.
We can relate the difference in charged particle density to the mass of the disrupted object as
\begin{equation}\label{eqn:mass}
    m = c\,\Delta\rho\, R_{\rm pc}\, \Delta t,
\end{equation}
where $\Delta t$ is the duration the pulsar remains in the altered spin-down state.
Taking a pulsar mass of $1.4\,M_\odot$, a radius of 10~km, $\Delta\dot{\nu} \sim 2 \times 10^{-15}$\,s$^{-2}$ from \citet{Brook2014} and extending $\Delta t$ to 2005--2023, we find a $\sim 3 \times 10^{15}$\,g object is needed to sustain the spin-down rate.
This is only a factor of six larger than the original estimate.
If the second, 2015 event was also due to an asteroid interaction then the change in spin-down of $\Delta\dot{\nu} \sim 0.4\times 10^{-15}$\,s$^{-2}$ would correspond to a change in charged particle density of $\sim 2 \times 10^{-9}$\,C\,cm$^{-3}$.
Combined with the (approximately) $7$\,yr duration, the 2015 event is then consistent with the disruption of a smaller $\sim 2 \times 10^{14}$\,g object.
Both our revised 2005 asteroid mass estimate and that of the 2015 event are well within the mass-range of known small Solar System bodies as well as expectations for debris disks orbiting pulsars \citep{Cordes2008}.

While the asteroid-interaction hypothesis provides a plausible explanation for the 2005 and 2015 events, the precursor component being active during observations taken in the 1970s/80s and after the 2005 event presents somewhat of a complicating factor.
If its appearance is triggered by the disruption of minor bodies falling toward the pulsar, then its presence in these early observations would imply there is a preferential region of the magnetosphere that the ionised debris is funnelled into.
Smaller profile shape changes that occurred in 1992 and 2010 were also not accompanied by detectable changes in $\dot{\nu}$.
However, the short-timescale wobbles in $\dot{\nu}$ seen in Figures~\ref{fig:long_var} and \ref{fig:pol_var} could have masked associated spin-down variations that were smaller than $\sim 10^{-16}$\,s$^{-2}$.

\subsection{Comparison with other state-changing pulsars}

Correlated emission/spin-down state switching pulsars hereto identified can be broadly categorised into three distinct populations: i) those where the variations occur on some quasi-periodic time-scale; ii) pulsars that undergo transient variations that recover over some finite timescale; iii) and those that remain in the altered state for an extended period.
It is useful to compare pulsars within each category, as similarities in their intrinsic properties and observed behaviour can provide clues as to what causes these variations.
From Figures~\ref{fig:long_var} and \ref{fig:pol_var} it is clear that PSR~J0738$-$4042 falls into the latter category, exhibiting both long-term and transient changes in its profile shape and spin-down rate.

The September 2005 event bears a striking resemblance to a sustained profile/spin-down event that occurred in PSR~J2037$+$3621 in early 2003 \citep{Shaw2022}.
This pulsar underwent an initial transient increase in $|\dot{\nu}|$ before settling into a sustained high-$|\dot{\nu}|$ state similar to that of our $|\dot{\nu}|$ timeseries shown in Figure~\ref{fig:long_var} but having been flipped about the vertical axis.
This event in PSR~J2037$+$3621 was not associated with the emergence of a new profile component, but rather a sudden drop in emission from its leading and trailing components and an enhancement of the central peak.
In contrast, the transient December 2015 event in PSR~J0738$-$4042 was similar to variations seen in PSRs J1602$-$5100 and J2043$+$2740 \citep{Brook2016, Shaw2022}. 
Both had new, transient components that appeared in their pulse profiles and displayed significant deviations in $|\dot{\nu}|$ that recovered back to their pre-event states after several hundred days.
PSR~J1602$-$5100 exhibited a reduction in $|\dot{\nu}|$ associated with transient decrease in profile main peak, increased emission from two trailing components, whereas PSR~J2043$+$2740 has displayed two separate increases in spin-down rate that were each associated with the enhanced emission from a component on the profile trailing edge. 
Unlike the transient events in PSRs~J0738$-$4042 and J1602$-$5100, the profile shape-change associated with the second event in PSR~J2043$+$2740 lagged the ramp-up in $|\dot{\nu}|$ by $\sim 400$\,d. 
This event was also preceded by an additional, short-term increase in $|\dot{\nu}|$ that may not have been associated with a profile shape change.
However, \cite{Shaw2022} note the lack of observations during this short-term event makes this somewhat uncertain.
If we assume the events in these three other pulsars were all caused by in-falling asteroids, using Equations~\ref{eqn:charge} and \ref{eqn:mass} to relate their changes in $|\dot{\nu}|$ with event duration, we find the corresponding object masses range between $\sim 10^{14}$--$10^{15}$\,g.
This is consistent with the aforementioned mass estimates for the PSR~J0738$-$4042 events.

Both transient and sustained changes in profile shape and spin-down rate are observed in radio-loud magnetars and high-magnetic field strength pulsars \citep[see][]{Camilo2008, Scholz2017, Dai2018, Lower2021b}. 
These phenomena are typically associated with impulsive high-energy outbursts that are believed to be triggered by magnetic reconnection events and either crustal fracturing or plastic motion in these objects \citep{Duncan1992, Thompson2002, Younes2022}.
PSR~J0738$-$4042 has displayed similar pulse-profile changes to those seen in magnetars. 
In particular, the appearance and disappearance of new profile components as well as drifting in pulse phase \citep{Levin2019, Lower2021a}.
However, there are several key differences between PSR~J0738$-$4042 and these highly magnetized objects that preclude a common origin to their observed radiative and rotational behaviour.
Magnetar outbursts result in rapid increases in their spin-down rates due to increased twists in their magnetic fields and enhanced particle outflows \citep{Beloborodov2009, Harding1999}. 
They are also prolific sources of high-energy radiation and often emit intense hard X-ray and gamma-ray bursts at the onset of an outburst \citep{Duncan1992}, and often display highly unstable linear polarisation position angle swings due to their dynamic magnetic fields \citep{Levin2012, Dai2019, Lower2021a}.
The most comparable profile/spin-down event in PSR~J0738$-$4042 resulted in a sustained decrease in spin-down rate, as opposed to a rapid increase.
Additionally, neither of the 2005 or 2015 events in PSR~J0738$-$4042 were reported to have been associated with a high-energy transient, nor resulted in a substantial alteration of the linear polarisation position angle swing.

\section{Conclusion}\label{sec:conclusion}

We have undertaken a comprehensive analysis of the polarised radio emission and rotational properties of PSR~J0738$-$4042 using data collected from four different radio observatories.
RVM fits to the linear polarisation position angle swing across the profile, combined with the measured profile width suggest the pulsar is a near-aligned rotator with $\alpha = 160\degr$ and a radio emission height of $330$\,km. 
Our updated timing measurements imply extremely large braking indices that depend on whether the pre-/post-2005 or the full timing baselines are analysed, indicating the non-stationary timing variations in the pulsar can be absorbed into the inferred value of $\ddot{\nu}$.
Indeed, performing a fit where the timing data around the 2005 event were ignored returned a value of $\ddot{\nu}$ that is consistent with zero, and improved measurements of the intrinsic pulsar parameters.
We rule out a previously reported glitch during the December 2015 event was ruled out \citep{Zhou2023}, indicating the correlated profile/spin-down behaviour in this pulsar are more likely to be magnetospheric in origin.
We argue the smaller pre-2005 braking index is unlikely to have resulted from the gravitational influence of a binary companion, though the presence of a $\lesssim 0.24$\,$M_{\odot}$ star or more massive compact-object cannot be ruled out.
The updated pulsar proper motion from our fit to the full timing dataset is mid-way between previous pulsar timing measurements \citep{Downs1983} and that obtained from interferometric observations from the VLA \citep{Brisken2003}.
This tension motivates future measurements of a parallax distance and improved proper-motion using very-long baseline interferometry.

Gaussian process regression was used to construct a continuous model for the temporal evolution of $\dot{\nu}$ over 50\,yr.
In addition to reproducing previously identified spin-down deviations \citep{Brook2014, Zhou2023}, we found the previous disappearance of a transient precursor component in the radio profile between 1981--1988 was not associated with a substantial change in $\dot{\nu}$. 
The lack of a corresponding spin-down event presents a curious departure from the known behaviour of this pulsar, and may be evidence that the correlated emission/rotation events are not the result of the pulsar switching between discrete states.
We also employed Gaussian process regression to model the profile variations in the pulsar between 2008--2023.
Changes in both the linear and circular polarisation components largely track the total intensity profile.
Multi-band variations in observations taken over four years with the Parkes-UWL show similar profile changes are a wideband phenomenon.
Apparent differences between subbands are the result of frequency-dependent profile evolution.

Behaviour similar to that observed in PSR~J0738-4042 has been detected among a growing sample of pulsars \citep{Lyne2010, Brook2016, Shaw2022}. 
It is clear that pulsars exhibit stochastic variations in their timing and their profiles over timescales from ms to decades. These variations have multiple possible origins ranging from fluctuations in the neutron-star interior, magnetic field structure and plasma density, through to structures in the circumstellar and interstellar medium. 
Magnetospheric origins for this behaviour are favoured, however the exact triggering mechanism is unknown.
Here, we revisited the hypothesis of an asteroid interaction as a potential trigger for the profile/spin-down variations.
The lack of a substantial recovery from the September 2005 event means the putative minor body that was disrupted had to be a factor of 10 more massive than initially estimated. 
The approximately 7-year duration of the December 2015 event can be attributed to a smaller, $\sim 10^{14}$\,g object.
Both are consistent with theoretical predictions for the masses of asteroids orbiting pulsars and estimates for similar transient events seen in three other radio pulsars \citep{Brook2016, Shaw2022}.
If the events are indeed due to asteroid interactions, then the high spacial velocity of the pulsar means they would have originated from a fall-back disk or were dynamically captured.
Combined with the challenges associated with reconciling the precursor disappearance in the 1980s with expectations from magnetospheric state switching, the asteroid hypothesis remains a possible explanation for the variability in PSR~J0738$-$4042.
Future observations taken at infrared or sub-mm wavelengths could confirm or rule out a the existence of a debris disk around the pulsar.

\section*{Acknowledgements}

We thank Lawrence Toomey for help with extracting the archival Deep Space Network pulse arrival times.
We also thank Andrew Lyne for insightful comments on the manuscript.
The Parkes radio telescope (\textsl{Murriyang}) is part of the Australia Telescope National Facility (\href{https://ror.org/05qajvd42}{https://ror.org/05qajvd42}) which is funded by the Australian Government for operation as a National Facility managed by CSIRO.
The Molonglo Observatory is owned and operated by the University of Sydney.
Major support for the UTMOST project has been provided by Swinburne University of Technology and the Mt. Cuba Astronomical Foundation.
We acknowledge the Wiradjuri and Ngunnawal people as the respective traditional custodians of the Parkes and Molonglo Observatory sites.
This work was supported by resources and expertise provided by CSIRO IMT Scientific Computing.
Some of the data analysis behind this work was performed using the OzSTAR national HPC facility, which is funded by Swinburne University of Technology and the National Collaborative Research Infrastructure Strategy (NCRIS). 
PRB is supported by the UK Science and Technology Facilities Council (STFC), grant number ST/W000946/1.
RMS acknowledges support through Australian Research Council (ARC) Future Fellowship FT190100155. 
Part of the work was undertaken with support from the ARC Centre of Excellence for Gravitational Wave Discovery (CE1700004).
Work at NRL is supported by NASA.
Pulsar research at the Jodrell Bank Centre for Astrophysics is supported by a consolidated grant from the UK STFC.
This work has made use of NASA's Astrophysics Data System.

\section*{Data Availability}

The archival Deep Space Network timing data are available from \citet{Downs1983} and \citet{Downs1986}.
Parkes P574 data are available to download via the CSIRO Data Access Portal (\href{https://data.csiro.au/}{https://data.csiro.au/}) following an 18-month proprietary period.
Other data and data products are available upon reasonable request to the corresponding author.



\bibliographystyle{mnras}
\bibliography{main} 




\appendix


\bsp	
\label{lastpage}
\end{document}